\documentclass[sigconf,nonacm]{acmart}
\acmConference[MSR '26]{Mining Software Repositories}{April 13--14, 2026}{Rio de Janeiro, Brazil}
\acmYear{2026}
\copyrightyear{2026}

\acmBooktitle{Mining Software Repositories (MSR '26), April 13--14, 2026, Rio de Janeiro, Brazil}
\acmPrice{15.00}
\acmISBN{978-x-xxxx-xxxx-x/26/04}
\acmDOI{10.1145/xxxxxxxxx.xxxxxxx}

\usepackage[dvipsnames,table]{xcolor}
\usepackage{multicol}
\usepackage{enumitem}
\usepackage{amsmath}
\usepackage{listings}
\usepackage{framed}
\setlist[itemize]{leftmargin=1em}
\usepackage{array}
\newcolumntype{C}[1]{>{\centering\arraybackslash}p{#1}}
\usepackage[ruled,vlined]{algorithm2e}

\usepackage{subcaption}
\usepackage{booktabs}
\usepackage{multirow}
\usepackage{enumitem}
\usepackage[table]{xcolor}
\setlist[itemize]{leftmargin=*}
\setlist[enumerate]{leftmargin=*}

\definecolor{niceblue}{HTML}{0000FF}
\definecolor{blue}{HTML}{000000}

\newcommand{\here}[1]{%
  \hypertarget{resp:#1}{}
  \fcolorbox{niceblue}{niceblue!15}{%
    \label{resp:#1}%
    \bfseries\scriptsize  {#1}%
  }%
}
\renewcommand{\here}[1]{}

\keywords{Causality, Correlation, Software Analytics, Explanation, Stability, Optimization}

\title{Can Causality Cure Confusion Caused By Correlation (in Software Analytics)?}

\author{Amirali Rayegan}
\email{arayega@ncsu.edu}
\affiliation{%
  \institution{North Carolina State University}
  \city{Raleigh}
  \state{North Carolina}
  \country{USA}
}

\author{Tim Menzies}
\email{tjmenzie@ncsu.edu}
\affiliation{%
  \institution{North Carolina State University}
  \city{Raleigh}
  \state{North Carolina}
  \country{USA}
}

\begin{document}

\begin{abstract}

\textbf{Background:} Symbolic models, particularly decision trees, are widely used in software engineering for explainable analytics in defect prediction, configuration tuning, and software quality assessment. Most of these models rely on correlational split criteria, such as variance reduction or information gain, which identify statistical associations but cannot imply causation between X and Y. Recent empirical studies in software engineering show that both correlational models and causal discovery algorithms suffer from pronounced instability. This instability arises from two complementary issues: (1) Correlation-based methods conflate association with causation. (2) Causal discovery algorithms rely on heuristic approximations to cope with the NP-hard nature of structure learning, causing their inferred graphs to vary widely under minor input perturbations. Together, these issues undermine trust, reproducibility, and the reliability of explanations in real-world SE tasks.

\textbf{Objective:} This study investigates whether incorporating causality-aware split criteria into symbolic models can improve their stability and robustness, and whether such gains come at the cost of predictive or optimization performance. We additionally examine how the stability of human expert judgments compares to that of automated models.

\textbf{Method:} 
Using 120+ multi-objective optimization tasks from MOOT\footnote{tiny.cc/moot} repository of multi-objective optimization tasks, we evaluate stability through a preregistered bootstrap-ensemble protocol that measures variance with win-score assignments. We compare stability of humans causal assessments with correlation-based decision trees (EZR). We would also compare the causality-aware trees which leverage conditional-entropy split criteria and confounder filtering. Stability and performance differences are analyzed using statistical methods (\color{blue}variance, Gini Impurity, \color{black}KS test, Cliff’s delta).

Note that all human studies conducted as part of this study will comply with the policies of our local Institutional Review Board.

\end{abstract}

\maketitle

\section{Introduction}
Symbolic models such as decision trees and rule-based learners are central to software analytics because they provide the rare combination of predictive power and human interpretability. They are widely used to explain configuration optimizations, guide quality assessments, and support defect prediction, where practitioners need transparent reasoning rather than opaque statistical correlations. Many symbolic models used in software engineering today rely on correlational split criteria (variance reduction, information gain, Gini impurity). These measures do not know if X truly causes changes in Y, whether Y drives X, or whether a third variable Z confounds both. Recent work has exposed the consequences of this limitation. In the Shaky Structures study\cite{hulse2025shaky}, causal discovery algorithms were shown to be highly unstable on real software engineering data: minor perturbations in sampling or preprocessing lead to radically different causal graphs, raising concerns about reliability and reproducibility. At the same time, correlation-based models, though easier to train, are sensitive to noise and can produce explanations that fluctuate widely under bootstrapping or small dataset shifts. Together, these findings suggest that both sides of the current methodological divide (correlational trees and causal discovery tools) suffer from instability which undermines practitioner trust in data-driven conclusions.

\begin{figure*}[ht!]
\centering
\scriptsize
\begin{framed}
\begin{lstlisting}
if TIME <= 4                 >       if PCON > 4                 >       if PVOL > 3                >       if ACAP > 2
|  if STOR <= 4              >       |  if TOOL <= 3;            >       |  if FLEx <= 2;           >       |  if LTEx > 3
|  |  if ACAP > 3;           >       |  if TOOL > 3;             >       |  if FLEx > 2             >       |  |  if DATA > 4;
|  |  if ACAP <= 3           >       if PCON <= 4                >       |  |  if TIME <= 4         >       |  |  if DATA <= 4
|  |  |  if TOOL > 3;        >       |  if TEAM > 4              >       |  |  |  if TOOL <= 3;     >       |  |  |  if PMAT <= 3;
|  |  |  if TOOL <= 3;       >       |  |  if PCAP > 3;          >       |  |  |  if TOOL > 3;      >       |  |  |  if PMAT > 3;
|  if STOR > 4;              >       |  |  if PCAP <= 3          >       |  |  if TIME > 4          >       |  if LTEx <= 3
if TIME > 4                  >       |  |  |  if PREC <= 5;      >       |  |  |  if PCAP > 2;      >       |  |  if CPLx > 5;
|  if RUSE > 4;              >       |  |  |  if PREC > 5;       >       |  |  |  if PCAP <= 2;     >       |  |  if CPLx <= 5
|  if RUSE <= 4              >       |  if TEAM <= 4             >       if PVOL <= 3               >       |  |  |  if ARCH <= 3;
|  |  if TEAM > 3;           >       |  |  if PCAP > 3           >       |  if LTEx <= 3            >       |  |  |  if ARCH > 3
|  |  if TEAM <= 3           >       |  |  |  if TOOL <= 3;      >       |  |  if SITE <= 3;        >       |  |  |  |  if SITE <= 1;
|  |  |  if ACAP > 3;        >       |  |  |  if TOOL > 3;       >       |  |  if SITE > 3;         >       |  |  |  |  if SITE > 1;
|  |  |  if ACAP <= 3;       >       |  |  if PCAP <= 3;         >       |  if LTEx > 3;            >       if ACAP <= 2;
\end{lstlisting}
\end{framed}
\caption{Instability of correlation-based splitting. Using EZR on the same MOOT dataset (coc1000.csv), four decision trees trained with different random seeds produce markedly different structures, feature selections, and explanations. Demonstrating how correlational split criteria can yield highly inconsistent models even under identical data.}
\label{fig:instability_example}
\end{figure*}

Despite this instability, symbolic models are often used for understanding optimization behavior in software systems. The EZR framework\cite{menzies2025compact}, introduced in the Minimal Data, Maximum Clarity study\cite{rayegan2025minimal}, demonstrated that small, actively sampled decision trees can produce succinct, actionable explanations while achieving near–state-of-the-art optimization performance across dozens of datasets. However, EZR and most interpretable learners in SE still builds its explanations on associational split criteria. As illustrated in Figure~\ref{fig:instability_example}, even correlation-based trees built by EZR on the same dataset can diverge substantially when only the random seed is changed.

This creates a critical open question: would symbolic models become more stable, interpretable, and robust if their split criteria were grounded in causal information rather than correlation? Addressing this question has immediate practical value for software engineering research and practice. If causality-aware splitting can reduce explanation divergence, improve prediction consistency, and identify more coherent feature sets, it may offer a new pathway toward reliable, human-centered analytics in a field increasingly concerned with trust, reproducibility, and actionable insight.

\section{Motivation and related work}
Causality has become an increasingly important theme in software engineering, motivated by the desire to move beyond associations and toward deeper mechanistic understanding of how software systems behave. As Pearl notes, causality allows analysts to move from “seeing” correlations to “doing” interventions and “imagining” counterfactuals\cite{pearl2018book}. Within SE, causal graphs are widely used for tasks such as defect prediction, configuration tuning, test failure analysis, and project management\cite{chadbourne2023applications, giamattei2025causal, giamattei2024causality, johnson2020causal}. However, recent evidence highlights severe instability in causal graphs generated from SE data. Hulse et al. study\cite{hulse2025shaky} systematically demonstrates that causal discovery algorithms (PC, FCI, GES, LiNGAM~\cite{Spirtes2000, Spirtes2002, Chickering2002, Shimizu2006}) often produce wildly different graphs when applied to small perturbations in data, tuning parameters, or algorithmic choices(across projects, releases, bootstrap samples, and generators) and even more than half of all causal edges typically change. They warn that such instability undermines confirmatory, generative, and post-hoc causal studies, and cautions that naively trusting causal graphs in SE can lead to inconsistent or reversed conclusions (For example “LOC $\rightarrow$ bugs” in one release vs. “bugs $\rightarrow$ LOC” in another). 
For other examples of this instability, see section 2.1 of \cite{hulse2025shaky}.

In contrast to causal approaches, most software analytics relies on correlation-based learners(decision trees, random forests, regression models, and similarity heuristics) which operate solely on Pearl’s associational rung of ladder of causation~\cite{pearl2018book}. These models cannot distinguish why a correlation appears; whether it's

\begin{enumerate}
    \item \textbf{direct causation}, $X\rightarrow Y$ meaning changes in X genuinely influence Y; or
    \item \textbf{reverse causation}, $Y\rightarrow X$ where the apparent effect flows backward; or
    \item \textbf{confounding}, $X \leftarrow Z\rightarrow Y$ where a third factor Z creates a spurious association.
\end{enumerate}
The Shaky Structures study highlights this ambiguity and shows that correlation-based rules frequently capture relationships that are not causally meaningful~\cite{hulse2025shaky}. Because common split criteria (variance reduction, information gain, and impurity minimization) treat all associations as equally valid, symbolic models often elevate spurious and confounded features to prominent decision rules, making them fragile under small data perturbations. This instability is widely documented in SE, where minor sampling changes can lead to significantly different trees, feature rankings, or optimization explanations \cite{lustosa2024isneak,senthilkumar2024can,rayegan2025minimal}(Figure~\ref{fig:instability_example}). Consequently, while computationally attractive, correlational methods remain vulnerable to instability driven by sampling variance and hidden confounders, limiting their reliability for decision-making and explanation in SE practice.

\section{Research Questions}
In this study, we assess model instability using a variance-based metric (Section 5) that measures how consistently ensembles of symbolic models assign win scores to the same test instances under bootstrap perturbations. Using this metric, we investigate the following research questions:
\begin{enumerate}
    \item \textbf{RQ1 — Human vs. Automated Stability:} When humans are asked to judge causal relationships within a domain, do they reach stable conclusions? And when we apply our instability metric to human-labeled causal assessments, how does this stability compare to correlation-based symbolic models?
    \color{blue}
    \item \textbf{RQ2 — Effect of Causal Reasoning on Model Performance Stability:} When using automated learning methods, does incorporating causality-aware split criteria produce more stable results than traditional correlation-based criteria?
    \color{black}
    \item \textbf{RQ3 — Trade-offs Between Stability and Performance:} 
    If causal reasoning improves model stability, does it introduce any performance penalties in prediction or optimization? Specifically, does potential model stability improvement through causal reasoning lead to reductions in predictive or optimization performance?
\end{enumerate}
\begin{table*}[t]
\scriptsize
\renewcommand{\baselinestretch}{0.9}
\caption{Summary of the datasets used in this study.}
\label{datasets-summary}
\begin{tabular}{@{}C{1cm}@{~}p{3cm}p{4.5cm}p{5cm}p{1cm}p{1.2cm}@{}}
\# \textbf{Datasets} & \textbf{\qquad \quad Dataset Type}  & \textbf{File Names} & \textbf{Primary Objective}        & \textbf{x/y}          & \textbf{\# Rows}       \\ \midrule
25          & \begin{tabular}[c]{@{}l@{}}Specific Software Configurations\end{tabular} & SS-A to SS-X, billing10k                                                 & Optimize software system settings                                     & 3-88/2-3   & 197–86,059   \\ 
12          & \begin{tabular}[c]{@{}l@{}}PromiseTune Software\\Configurations\end{tabular} & \begin{tabular}[c]{@{}l@{}} 7z, BDBC, HSQLDB, LLVM, PostgreSQL, dconvert, \\deeparch, exastencils, javagc, redis, storm, x264\end{tabular}                                                 & Software performance optimization                                     &  9-35/1  &  864-166,975  \\ \midrule
1           & Cloud                            & HSMGP num                                                                & Hazardous Software, Management Program data                           & 14/1         & 3,457     \\
1           & Cloud                            & Apache AllMeasurements                                                   & Apache server performance optimization                                & 9/1          & 192     \\
1           & Cloud                            & SQL AllMeasurements                                                      & SQL database tuning                                                   & 39/1         & 4,654    \\
1           & Cloud                            & X264 AllMeasurements                                                     & Video encoding optimization                                           & 16/1         & 1,153 \\
7           & Cloud                            & (rs—sol—wc)*                                                             & misc configuration tasks                                              & 3-6/1      & 196–3,840 \\ \midrule
35          & Software Project Health          & Health-ClosedIssues, -PRs, -Commits                                      & Predict project health and developer activity                         & 5/2-3      & 10,001  \\ \midrule
3           & Scrum                            & Scrum1k, Scrum10k, Scrum100k                                             & Configurations of the scrum feature model                             & 124/3      & 1,001–100,001  \\ \midrule
8           & Feature Models                   & FFM-*, FM-*                                                              & Optimize number of variables, constraints and Clause/Constraint ratio & 128-1,044/3 & 10,001 \\ \midrule
1 &	Software Process Model &	nasa93dem &	Optimize effort, defects, time and LOC	& 24/3 &	93  \\
1           & Software Process Model           & COC1000                                                                  & Optimize risk, effort, analyst experience, etc                        & 20/5         & 1,001   \\
4           & Software Process Model           & POM3 (A–D)                                                               & Balancing idle rates, completion rates and cost                       & 9/3          & 501–20,001 \\
4           & Software Process Model    & XOMO (Flight, Ground, OSP)                                               & Optimizing risk, effort, defects, and time                            & 27/4         & 10,001  \\ \midrule
3           & Miscellaneous                             & auto93, Car\_price, Wine\_quality                                        & Miscellaneous                                                         & 5-38/2-5   & 205–1,600 \\ \midrule
4           & Behavioral                       & all\_players, student\_dropout,\newline HR-employeeAttrition, player\_statistics & Analyze and predict behavioral patterns                              & 26-55/1-3  & 82–17,738  \\ \midrule
4           & Financial                        & BankChurners, home\_data, Loan, Telco-Churn                              & Financial analysis and prediction                                     & 19-77/2-5  & 1,460–20,000 \\ \midrule
3           & Human Health Data                & COVID19, Life\_Expectancy, hospital\_Readmissions                        & Health-related analysis and prediction                                & 20-64/1-3  & 2,938–25,000    \\ \midrule
2           & Reinforcement Learning           & A2C\_Acrobot, A2C\_CartPole                                              & Reinforcement learning tasks                                          & 9-11/3-4   & 224–318       \\ \midrule
5           & Sales                            & accessories, dress-up, Marketing\_Analytics, \newline socks, wallpaper            & Sales analysis and prediction                                         & 14-31/1-8  & 247–2,206 \\ \midrule
2	& Software testing	& test120, test600	& Optimize the class	& 9/1	& 5,161\\ \midrule

\textbf{127}         & \textbf{Total} & & &              &            
\end{tabular}
\end{table*}

\section{Method}

\subsection{Baseline Method: EZR a Correlation-Based Splitting}
This study builds on the EZR framework \cite{menzies2025compact}, a lightweight and label-efficient reasoning system designed to generate small, interpretable decision trees for multi-objective optimization tasks. EZR begins by selectively sampling the dataset, labeling only a small number of instances, and estimating their quality using the \textit{win score} of each row’s \textit{distance-to-heaven (d2h)}. The d2h metric provides a scalar measure of solution quality by computing the Euclidean distance between a row’s normalized objective values and the ideal “heaven” point, where all objectives achieve their best possible values. The win score then maps the predicted d2h into a normalized improvement scale (0–100), assigning higher values to data points closer to ideal and enabling performance comparisons across datasets.

Using these quality estimates, EZR constructs a symbolic decision tree model by recursively splitting the data. At each internal node, it evaluates candidate attributes and selects the split that yields the greatest reduction in variance, a classical decision-tree heuristic that partitions the data into increasingly homogeneous subsets. This correlation-based approach produces concise, human-readable rules that achieve strong predictive performance while retaining interpretability.

While EZR provides a strong baseline and has been shown to generate compact, useful explanations across a wide range of optimization tasks\cite{rayegan2025minimal}, its split criteria are fundamentally correlational. That is, the algorithm selects attributes that statistically associate with the outcomes, without distinguishing between genuine causal drivers and coincidental or confounded relationships.

\subsection{Causality-Aware Splitting via Conditional Entropy}
To operationalize causal reasoning within the symbolic model, this study uses a causal module that uses a set of causality-inspired split criteria. Instead of measuring how much an attribute reduces variance, the causal splitter evaluates each candidate attribute X by computing how well it explains the variability in the target Y through \textbf{conditional entropy}. As shown in equation~\ref{eq:cond_ent} conditional entropy measures the remaining uncertainty in Y after observing X. 

\begin{equation}
H(Y \mid X) = -\sum_{x} p(x) \sum_{y} p(y \mid x) \log_2 p(y \mid x)
\label{eq:cond_ent}
\end{equation}

A low value of $H(Y\mid X)$ indicates that knowledge of X substantially reduces uncertainty in Y, consistent with the idea that X may have a directional influence on Y. Next, to normalize across datasets and ensure comparability, the causal splitter evaluates the fraction of Y’s entropy that remains unexplained given X. Equation~\ref{eq:causalScore} shows how the normalization happens.

\begin{equation}
    \text{CausalScore}(X) = \frac{H(Y \mid X)}{H(Y)}
    \label{eq:causalScore}
\end{equation}

Lower scores indicate attributes that better “explain” Y and are more plausible causal candidates. The causal discretizes numerical values, handles symbolic attributes, and evaluates candidate splits(all possible feature/value pairs) by selecting those with the smallest causal scores.
\begin{table*}[t]
\color{blue}
\centering
\small
\renewcommand{\arraystretch}{1.05}
\caption{\color{blue}Comparison of Decision Tree Variants}
\label{tab:dt-compare}
\begin{tabular}{@{}p{2.5cm} p{4.5cm} p{4.0cm} p{4.5cm}@{}}
 & \textbf{C4.5} & \textbf{ID3} & \textbf{Ours} \\
\midrule
\rowcolor{blue!5}
\textbf{Entropy of data} &
$\displaystyle H(y) = - \sum_i P_i \log_2 P_i$ &
$\displaystyle H(y) = - \sum_i P_i \log_2 P_i$ &
$\displaystyle H(y) = - \sum_i P_i \log_2 P_i$ \\

\textbf{Finding splits} &
Highest info.\ Gain ratio &
Highest info.\ gain &
Lowest Conditional Entropy \\
\rowcolor{blue!5}
\textbf{Split formula} &
\raggedright
$\displaystyle
\frac{\text{info.\ gain}}{\text{SplitInfo}}
=
\frac{ H(y) - \sum_{\nu} \frac{|Y_{\nu}|}{|Y|} H(y_{\nu}) }
      { - \sum_{\nu} \frac{|Y_{\nu}|}{|Y|} \log_2 \frac{|Y_{\nu}|}{|Y|} }
$ &
\raggedright
$\displaystyle
H(y) - \sum_{\nu} \frac{|Y_{\nu}|}{|Y|} H(y_{\nu})
$ &
$\displaystyle
\frac{ H(y|x) }{ H(y) }
=
\frac{ - \sum_{i,\nu} p(x,y)\log_2 p(x) }
     { - \sum_i P_i \log_2 P_i }
$ \\
\textbf{Pruning} &
Post-pruning (subtree simplification) &
--- &
Pre-pruning (removing confounders)\\
\rowcolor{blue!5}
\textbf{Objective} &
Single &
Single &
Multi objective \\

\textbf{Data compatibility} &
Categorical and numerical &
Categorical &
Categorical and numerical \\
\end{tabular}
\end{table*}
\color{blue}
\here{B.1}
Our causality-aware split criterion is closely related to classical entropy-based decision tree learners such as ID3\cite{quinlan1987decision} and C4.5\cite{quinlan2014c4}. While minimizing conditional entropy $H(Y\mid X)$ (or its normalized form $H(Y\mid X)/H(Y)$) is mathematically equivalent to maximizing information gain used in these methods, our approach differs in three important ways (summarized in Table~\ref{tab:dt-compare}). First, we explicitly compute the normalized conditional entropy ratio to support clearer causal interpretation rather than classification purity. Second, instead of relying on post-pruning heuristics, we apply pre-pruning by removing confounders whose associations disappear under conditioning. Third, our trees are optimized for multi-objective optimization rather than single-objective classification. Consequently, although our split criterion shares theoretical roots with ID3 and C4.5, the intended use and resulting models differ substantially. Accordingly, our empirical baseline is EZR’s variance-reduction splitting rather than information-gain-based trees.

\color{black}

\subsection{Backdoor Protection: Confounder Filtering}
To handle potential confounding effects we also perform conditional mutual information checks. Variables that behave like confounders(variables whose association with Y disappears when conditioning on other features) are automatically identified and removed from consideration. This step aims to eliminate cases where an apparent correlation between X and Y is more likely the result of $Z \rightarrow [X,Y]$. To detect confounders, we evaluate each feature’s marginal mutual information with the target Y. A variable X is flagged as a potential confounder if it shows substantial mutual dependence with Y. For every such X, we then test whether this dependence vanishes when conditioning on another feature Z by computing the conditional mutual information $I(X;Y\mid Z)$. If $I(X;Y\mid Z)$ falls below a small threshold, this indicates that the apparent association between X and Y is fully explained by Z, matching the backdoor pattern $Z \rightarrow [X,Y]$. Features meeting this criterion are treated as confounders and removed before tree construction.

At the end, the resulting decision tree is constructed using splits intended to retain only $X \rightarrow Y$ relationships while suppressing spurious or confounded associations. This produces a causal-tree model that is directly comparable to the variance-based EZR tree, allowing us to isolate and evaluate the impact of causal reasoning on stability, interpretability, and robustness.

\color{blue}
\here{B.2}
Conditional mutual information has been extensively studied for feature selection in machine learning \cite{wollstadt2023rigorous}. Our backdoor protection applies these well-established techniques, but with a specific causal interpretation. Rather than filtering for relevance or redundancy, we identify features matching Pearl's backdoor pattern $Z \rightarrow [X,Y]$, where associations are explained away by conditioning. While the mathematical tools are standard, our research question asks whether this causally motivated filtering improves model stability in SE optimization contexts, an empirical question distinct from methodological novelty in feature selection.

\here{B.3}
As a limit to keep in mind, it is worth mentioning that our filtering method cannot establish causal directionality. Conditional mutual information is symmetric and correlational. We apply a heuristic that removes fully-mediated associations, assuming these are more likely spurious, but cannot determine the true causal structure. ($Z \rightarrow [X,Y]$ or $X \rightarrow [Z,Y]$) RQ2 empirically tests whether this heuristic improves stability compared to unrestricted correlation-based methods.
\color{black}


\section{Datasets and Sources and Collection}

The experiments in this study use the MOOT repository\cite{menzies2025mootrepositorymultiobjectiveoptimization}, a large and diverse collection of over 120 multi-objective optimization tasks drawn from recent software engineering research. MOOT spans a wide range of practical domains, including software configuration, cloud performance tuning, project and process analytics, feature models, and behavioral, financial, and health-related systems, ensuring coverage of realistic and heterogeneous optimization scenarios. These tasks originate from studies published in major SE venues (
ICSE~\cite{chen2026promisetune, DBLP:conf/icse/WeberKSAS23,10172849,DBLP:conf/icse/HaZ19}, 
FSE~\cite{nair2017using,DBLP:conf/sigsoft/JamshidiVKS18}, 
TSE~\cite{chen2025accuracy,xia2020sequential,krishna2020whence,Chen19,krall2015gale}, 
EMSE~\cite{hulse2025shaky, peng2023veer,guo2018data}, 
MSR~\cite{nair18}, 
ASE~\cite{nair2018faster}, 
TOSEM~\cite{lustosa2024learning}, 
IST~\cite{chen2018beyond,fu2016tuning}), reinforcing their relevance to real-world practice. Table~\ref{datasets-summary} summarizes the datasets in the repository along with key characteristics and category information. This breadth makes MOOT particularly well-suited for evaluating the generalizability, stability, and robustness of both correlational and causal symbolic models across varied data regimes.

\section{Study Design}
This study examines whether causality-aware symbolic models yield more stable and reliable conclusions than traditional correlation-based decision trees. We evaluate stability using a preregistered bootstrap-ensemble protocol(which will be explained in section 7.1) and compare human causal assessments with correlation-based models across a fixed set of tasks. Specifically, \textbf{RQ1} assesses the stability of human judgments relative to automated correlational splits; \textbf{RQ2} tests whether incorporating causal split criteria improves model stability; and \textbf{RQ3} investigates whether increased stability introduces trade-offs in predictive or optimization performance. The following subsections describe the study design for each research question and, in the next section, the evaluation metrics applied.

\subsection{Design for RQ1 : Human vs. Automated Stability}
RQ1 examines whether human judgments about which features directly influence which objectives are themselves stable and how this stability compares to the stability of automated models. To answer this question, we will conduct a small survey with domain experts using a curated subset of MOOT tasks. \color{blue} \here{D.1.1} Given the large number and diversity of MOOT datasets, recruiting domain experts for every task would be impractical. Accordingly, we conduct the human study on a carefully selected, representative subset of datasets from different categories, shown in Table~\ref{datasets-subset}\color{black}. For each selected optimization task, experts will be shown the feature set and objectives, and asked to identify which features they believe have a direct causal influence on each objective. Each expert’s assessment produces a $feature\rightarrow objective$ mapping for that task. To measure the stability of human judgments, we compute the variance across experts for each proposed $feature\rightarrow objective$ relationship. This yields a distribution of agreement magnitudes reflecting how consistently humans perceive causal influence within the same domain.
We then compare this human-derived scores to the scores produced by automated models trained on the same tasks. For each dataset, we generate 20 correlation-based EZR trees and record the variability of their selected top features and split choices using the same variance-of-answers framework applied to human data. The final analysis compares domain experts and correlation-based models on the same stability measure discussed in section 7.1. This establishes whether human causal intuition is more or less stable than automated correlational heuristics, and whether they match closely to human-level consistency.

\begin{table}[t]
\centering
\color{blue}
\footnotesize
\renewcommand{\arraystretch}{0.95}
\caption{\color{blue}Subset of MOOT datasets used in the human study.}
\label{datasets-subset}
\begin{tabular}{@{}p{1.8cm} p{5.3cm} c c@{}}
 
\textbf{Dataset Name} & \textbf{Category} & \textbf{x} & \textbf{y} \\ 
\midrule
COC1000 & Software Process Model & 20 & 5 \\
BDBC & Database System Configuration Optimization & 12 & 1 \\
SS-H & Software System Setting Optimization & 4 & 2 \\
Apache\newline AllMeasurements & Cloud Server Configuration Optimization & 9 & 1 \\
 
\end{tabular}
\end{table}

\color{blue}
\subsection{Design for RQ2 — Effect of Causal Reasoning on Model Performance Stability}
\here{B.6.1} RQ2 evaluates whether the performance of decision trees constructed with causal split criteria are more stable than those constructed with traditional correlation-based splitting when applied to \textbf{optimization tasks} in the MOOT repository. Experiments for RQ2 and RQ3 will be executed on all datasets in Table~\ref{datasets-summary}. In each optimization task, the goal is to select the row from the test set that achieves the best multi-objective outcome according to the d2h metric. For each dataset, we use 50\% of the data as training and 50\% as testing, consistent across all treatments. Each model family (correlation-based and causality-based) builds its decision tree using only the training portion. A decision tree performs optimization by traversing from the root to a leaf node for every row in the test set, computing the predicted d2h value for each row, and then selecting the test row with the lowest predicted d2h (i.e., the one closest to the ideal multi-objective outcome). 

Both treatments operate under identical conditions, same training/test split, labeling budget, parameter settings, and search protocol, ensuring that any observed performance differences arise solely from the choice of split criterion. Keep in mind that for the causal-based decision trees, we will apply \textbf{Confounder Filtering} on the dataset before building the model. Performance is then evaluated by comparing the actual d2h values of the optimal row suggestions from each model, producing parallel performance distributions for correlation-based and causality-based trees.

For RQ2, we start with generating 20 correlation-based decision trees and 20 causality-based decision trees per each dataset. Next, we feed the same held-out test set to all decision trees and keep the result of each of them. This produces two distribution of optimization performance for every dataset, each containing 20 values. Stability of these performances is then evaluated using the metric and statistical framework described in Section 7.2.
\color{black}

\subsection{Design for RQ3 — Trade-offs Between Stability and Performance}
RQ3 evaluates whether the \textbf{optimization performance} of symbolic models that use causality-aware split criteria significantly differs from those that rely on traditional correlation-based splitting. 
\color{blue}
\here{B.6.2} Here, we follow an experimental protocol similar to RQ2 with two key differences. First, instead of training 40 trees on a single fixed train–test split, we perform 20 independent train–test splits and train two models per split: one correlation-based tree and one causality-based tree. Second, the focus shifts from measuring stability to comparing optimization performance. The resulting performance distributions are analyzed using the statistical framework described in Section 7.3 to determine whether one method achieves significantly better optimization accuracy than the other.
\color{black}

\section{Analysis Plan}
This section preregisters the evaluation criteria and statistical procedures used to analyze each research question. Each subsection corresponds directly to one RQ and defines the data used, how the measurements are computed, and the statistical tests applied. All statistical decisions are fixed prior to experimentation.

\subsection{Analysis Plan for RQ1 — Human vs. Correlation-Based Stability}
\subsubsection{Participants}
Subject-matter experts (SMEs) will be recruited following approval from the North Carolina State University Institutional Review Board (IRB). This study is expected to qualify as Minimal Risk, as participation consists solely of responding to a brief set of email-based questions regarding technical domain knowledge. We do not collect sensitive personal information, health data, demographic identifiers, or content involving vulnerable populations. All responses will be anonymized and used exclusively for constructing the human causal judgments required for RQ1, as described in Section 6.1.

\subsubsection{Recruitment}
The Computer Science Department at NC State University maintains long-standing collaborations with industry partners and researchers, many with over a decade of domain experience. To recruit subject-matter experts for this study, we will distribute a broad email call for participation to these established contacts and affiliated professional networks. Based on prior experience with SE expert-survey recruitment, we anticipate a 2–5\% response rate. To ensure adequate coverage across dataset categories, we plan to issue approximately 300–500 recruitment invitations.
\color{blue}
\here{A.3.1} Respondents will be asked to self-identify their primary areas of expertise, and tasks will then be assigned accordingly such that each expert evaluates only datasets aligned with their domain knowledge. Each task in the human study (available in Table~\ref{datasets-subset}) is expected to receive responses from approximately 15 to 20 independent experts, which is sufficient to estimate inter-expert variability while remaining feasible given the limited availability of qualified participants.
\color{black}

\subsubsection{SME Qualification and Verification}
To ensure high-quality domain judgments, participants must meet at least one of the following criteria: (1) primary authorship of a paper contributing a MOOT dataset, (2) active maintenance or core contribution to the underlying software system, or (3) demonstrated research expertise in areas such as configuration tuning, performance modeling, defect prediction, or software processes.
\color{blue}
\here{A.3.2} Participants will be asked to indicate which criterion they satisfy at the time of response, and their stated expertise will be reviewed by the study authors for consistency with the assigned dataset domain. Responses that do not meet these criteria, or whose expertise cannot be reasonably verified, will be excluded from analysis. To ensure reliable stability estimates, each dataset must receive at least 10 complete expert responses to be included in the RQ1 analysis. Datasets that do not meet this minimum will be excluded.
\color{black}

\subsubsection{Contingency for Dropouts}
We do not require experts to respond on a fixed schedule, and answers may be submitted at any time within the study window. In cases of inconsistent participation or dropout, we will retain only those responses that provide complete answers for each dataset. Any partial or incomplete dataset-level answers will be excluded. This ensures the reliability of the human-judgment stability analysis.

\subsubsection{Statistical Analysis}
For RQ1, we compare the stability of human domain experts with the stability of correlation-based models using a shared three-level ordinal judgment scale: no impact, mild impact, and certain impact. Each expert’s answer for a $feature \rightarrow objective$ relationship is encoded as an ordinal value {0,1,2}, corresponding to these three levels of perceived influence. For each $feature \rightarrow objective$ pair in each dataset, we compute the inter-expert variance of these ordinal judgments, producing a distribution of human stability scores.
\color{blue}
\here{C.2} It is worth mentioning that high levels of inter-expert disagreement are not treated as noise
or grounds for exclusion. Instead, disagreement is explicitly captured by the variance metric and analyzed as the primary signal of interest in RQ1.
\color{black}

To obtain a parallel measure for correlation-based models, we convert the model’s feature score into the same three-level scale. For each of the 20 bootstrap-trained EZR models, the impact of each feature on the objective is categorized as:
\begin{itemize}
    \item no impact (feature never appears in splits or explanations)
    \item mild impact (feature appears occasionally or in deeper nodes) 
    \item certain impact (feature appears consistently in top splits or model explanations).
\end{itemize}

\color{blue}
\here{B.5} To provide a stability measure that is well-defined for ordinal data and robust to small sample sizes, we compute the Gini impurity for each $\textit{feature} \rightarrow \textit{objective}$ pair. Given the distribution of ordinal judgments $\{0,1,2\}$ with class probabilities $p_k$, Gini impurity is defined as:

\begin{equation}
G = 1 - \sum_{k \in \{0,1,2\}} p_k^2 .\label{eq:gini}
\end{equation}

Lower impurity values indicate higher consistency, while higher values reflect greater disagreement. We compute this score for human expert judgments and for the distributions induced by the 20 bootstrap-trained models after mapping their outputs to the same three-level scale. Comparing the resulting Gini impurity distributions allows us to assess which source(humans or correlation-based models) exhibits greater stability in identifying feature–objective relationships.

\color{black}

\subsection{Analysis Plan for RQ2 — Stability Comparison}
\subsubsection{Evaluation}
\color{blue}
\here{B.6.3}
Each dataset yields two distributions of performance values, one for correlation-based trees and one for causality-based trees. Our goal is to measure the consistency of the performances in each distribution. 
\color{black}
\subsubsection{Statistical Analysis}
\color{blue}
To assess whether causality-aware split criteria yield greater stability, we compute and compare the variance of the resulting performance distributions. A statistically significant difference in variance indicates that incorporating causal split criteria has a measurable impact (positive or negative) on the stability of optimization performance.
\color{black}

\subsection{Analysis Plan for RQ3 — Performance Comparison}
\subsubsection{Evaluation}
For RQ3, we compare the optimization performance of correlation-based and causality-based trees by analyzing the d2h values of the data points each model selects as optimal from the test set. Each treatment is evaluated across 20 repeated runs to remove the effect of randomness, producing a distribution of performance scores for each method. Lower d2h values (and correspondingly higher win scores) indicate better optimization outcomes.
\subsubsection{Statistical Analysis}
\color{blue}
\here{B.6.4} To determine whether causality-aware split criteria reach better performance, we compare these distributions using a non-parametric framework that evaluates both distributional difference and effect magnitude. Distributional similarity is assessed using the Kolmogorov-Smirnov\cite {lilliefors1967kolmogorov} statistic, which measures the maximum gap between empirical CDFs, together with Cliff’s Delta to quantify effect size.
\begin{equation}
D_{n,m} = \sup_{x} \left| F_{n}(x) - G_{m}(x) \right|
\label{eq:K_S}
\end{equation}
Equation~\ref{eq:K_S} (Kolmogorov Smirnov test) gives the maximum vertical difference between the empirical CDFs of two samples, where $F_{n}(x)$ and $G_{m}(x)$ are the distributions being compared. Two distributions are treated as statistically indistinguishable only when the KS distance is below the 95\% confidence threshold and the effect size remains in the small/medium range. This analysis determines whether potential improvements in stability (RQ2 results) come with any meaningful trade-offs in optimization accuracy.
\color{black}

\color{blue}
Note that RQ3 does not aim to rank performance \textbf{across datasets}.
Instead, performance comparisons are conducted independently
within each dataset, where only two treatments are evaluated:
correlation-based and causality-based trees. In this paired setting,
distributional comparison using the Kolmogorov Smirnov test (Equation~\ref{eq:K_S}),
together with Cliff’s Delta, is sufficient to determine whether one
method meaningfully outperforms the other on that dataset.

\here{D.1.2} To summarize results across the MOOT repository, we report the
number of datasets in which each treatment performs better, worse,
or indistinguishably. We further analyze these outcomes by grouping
datasets according to characteristics such as domain category,
number of features, and objective dimensionality to derive insights
into whether either method exhibits systematic advantages under
specific data conditions.
\color{black}
\begin{acks}
We thank Professor Robert Feldt for his insightful discussions, generous brainstorming, and valuable intellectual guidance that strengthened this Registered Report.
\end{acks}

\bibliographystyle{ACM-Reference-Format}
\bibliography{references}

\newpage

\end{document}